\documentclass{llncs}

\usepackage{amsmath}
\usepackage{textcomp}
\usepackage{marvosym}
\usepackage{wasysym}
\usepackage{amssymb}
\usepackage{hyperref}

\usepackage{enumitem}
\setlist{noitemsep, topsep=7pt}
\let\oldenumerate\enumerate
\renewcommand{\enumerate}{\vspace{-\topsep}\oldenumerate}

\usepackage{listings}
\begin{document}

\makeatletter
\newcommand*{\bdiv}{%
  \nonscript\mskip-\medmuskip\mkern5mu%
  \mathbin{\operator@font div}\penalty900\mkern5mu%
  \nonscript\mskip-\medmuskip
}
\makeatother

\title{Declaratively solving Google Code Jam problems with Picat}

\author{Sergii Dymchenko \inst{1} \and Mariia Mykhailova \inst{2}}

\institute{Independent Researcher,\\
\email{sdymchenko@progopedia.com}
\and
Independent Researcher,\\
\email{michaylova@gmail.com}}

\maketitle

\begin{abstract}
In this paper we present several examples of solving algorithmic problems from the Google Code Jam programming contest
with Picat programming language using declarative techniques: constraint logic programming and tabled logic programming.
In some cases the use of Picat simplifies the implementation compared to conventional imperative programming languages,
while in others it allows to directly convert the problem statement into an efficiently solvable declarative problem specification without inventing an imperative algorithm.
\end{abstract}


\section{Introduction}
Google Code Jam\footnote{\url{https://code.google.com/codejam}} (GCJ) is one of the
biggest programming competitions in the world: almost 50,000 participants registered in 2014, and 25,462 of them solved at least one task.

GCJ competitors can use any freely available programming language or system (including Picat \footnote{\url{http://picat-lang.org/}} described in this paper).
We show examples of solving GCJ problems with Picat using constraint logic programming and tabled logic programming.

Picat is a new logic-based multi-paradigm programming language. Picat shares many features with Prolog, especially B-Prolog \cite{zhou2012bprolog},
but also has many distinct features: optional destructive assignments, functions in addition to predicates, explicit non-determinism, list comprehensions \cite{zhou2014picat}.

\subsection*{Picat implementation of TPK algorithm}

To give an idea of Picat syntax to a reader unfamiliar with the language we present an implementation of TPK algorithm.
TPK is an algorithm proposed by D.\,E.\,Knuth and L.\,T.\,Pardo  \cite{knuth1976early} used to show basic syntax of a programming language.
The algorithm allows user to input 11 real numbers ($a_0 \dots a_{10}$).
After that for $i = 10 \dots 0$ (in that order) the algorithm computes value $y = f(a_i)$, where $f(t) = \sqrt{|t|} + 5 {t}^3$, and outputs a pair $(i, y)$ if $y \leq 400$, or $(i,$ TOO LARGE$)$ otherwise.

\begin{lstlisting}[caption={TPK algorithm in Picat}]
f(T) = sqrt(abs(T)) + 5 * T**3.
main =>
    N = 11,
    As = to_array([read_real() : I in 1..N]),
    foreach (I in N..-1..1)
        Y = f(As[I]),
        if Y > 400 then
            printf("%w TOO LARGE\n", I - 1)
        else
            printf("%w %w\n", I - 1, Y)
        end
    end.
\end{lstlisting}

Line 1 defines a function to calculate the value of $f$ (a function in Picat is a special kind of a predicate that always succeeds with a return value).
Lines 2--12 define the \texttt{main} predicate.
Line 4 uses list comprehension to read 11 space-separated real numbers into array \texttt{As}.
Line 5 defines a header of \texttt{foreach} loop: \texttt{I} goes from 11 to 1 with the step -1 (in Picat array indices are 1-based).
Lines 6--11 calculate the value of y and print the result using an `if-then-else' construct.
\texttt{printf} is similar to the corresponding C language function; \texttt{\%w} can be seen as a ``wildcard'' control sequence to output values of different types.

\section{The Problems}
For this section we have chosen a set of GCJ problems from different years to demonstrate different useful aspects of Picat: constraint programming,
top-down dynamic programming with tabling, and the \texttt{planner} module.

\subsection*{Triangle Areas\footnote{Problem link: \url{http://goo.gl/enHWlq}}}

``Triangle Areas'' is a problem from the round 2 of GCJ 2008.
The problem gives integers $N$, $M$ and $A$ and asks to find any
triangle with vertices in integer points with coordinates $0 \leq x_i \leq N$
and $0 \leq y_i \leq M$ that has an area of $\frac{A}{2}$, or to decide that it does not exist.

``Triangle Areas'' is almost perfect for solving with constraint logic programming.
Variables are discrete, constraints are non-linear, and we are looking for any feasible solution.
To come up with an effective model we need to notice that one vertex of the triangle can be chosen arbitrarily.
With this observation, the most convenient way to calculate the doubled triangle area is to place one vertex in $(0,0)$; then $2S = |x_2y_3 - x_3y_2|$.
(The same formula can be used in an imperative solution.)

For this problem we present complete source code of the solution.
For subsequent problems we omit the \texttt{main} predicate to save space.

\begin{lstlisting}[caption={Complete Picat program for the ``Triangle Areas'' problem}]
import cp.
import util.
model(N, M, A, Points) =>
    [X2, X3] :: 0..N,
    [Y2, Y3] :: 0..M,
    A #= abs(X2 * Y3 - X3 * Y2),
    Points = [X2, Y2, X3, Y3].
do_case(Case_num, N, M, A) =>
    printf("Case #%w: ", Case_num),
    if model(N, M, A, Points), solve(Points) then
        printf("0 0 %s\n", join([to_string(V) : V in Points]))
    else
        println("IMPOSSIBLE")
    end.
main =>
    C = read_int(),
    foreach (Case_num in 1..C)
        N = read_int(), M = read_int(), A = read_int(),
        do_case(Case_num, N, M, A)
    end.
\end{lstlisting}

Lines 1--2 load Picat modules for constraint programming and utility functions.
Lines 3--7 define the model with input parameters \texttt{N}, \texttt{M}, \texttt{A} and a list of output parameters \texttt{[X2, Y2, X3, Y3]}.
\texttt{::} and \texttt{\#=} are from the `cp' module.
With \texttt{::} we define possible domains for \texttt{X2, X3, Y2, Y3} variables,
and \texttt{\#=} from `cp' constraints both left and right parts to be equal.
After model evaluation \texttt{X2, X3, Y2, Y3} variables will not necessarily be instantiated to concrete values, but they will have reduced domains with possible delayed constraints
and will be instantiated later with \texttt{solve}.

Lines 8--14 define the \texttt{do\_case} predicate to process a single input case. Line 9 outputs case number according to the problem specification.
Lines 10--14 are an `if-then-else' construct that outputs point coordinates if it is possible to satisfy our \texttt{model} predicate and solve
(assign concrete values from the domain to every variable) the constraint satisfaction problem, or ``IMPOSSIBLE'' otherwise.
Line 11 uses an interesting Picat feature -- list comprehension -- which is very similar to what Python and many other modern programming languages have.

Lines 15--20 define the \texttt{main} predicate that reads the number of test cases \texttt{C} and for each test case reads \texttt{N, M, A} parameters and executes \texttt{do\_case}.

This Picat program is very similar to our constraint programming solution in ECL\textsuperscript{i}PS\textsuperscript{e} CLP \cite{gcj-eclipse-arxiv}.

A possible imperative solution in a mainstream programming language requires a more in-depth analysis of the problem.
First observations will be the same. We will also note that it is impossible to find required triangle
if $A > M \times N$, and for $A = M \times N$ triangle $(0,0), (N,0), (0,M)$ is a valid answer.
Now, for $A < M \times N$ we can represent A as $M(A \bdiv M) + (A \bmod M)$, $0 < A \bdiv M < N, 0 < A \bmod M < M$.
If we match this representation with the area formula, we can see that points $(0, 0), (1, M)$, and $(- A \bdiv M, A \bmod M)$
form a triangle with area $\frac{A}{2}$. If we shift this triangle $A \bdiv M$ units in positive direction along the $x$ axis, we will get
a triangle $(A \bdiv M, 0), (A \bdiv M + 1, M), (0, A \bmod M)$ that will match all the requirements.

Arguably, our declarative solution in Picat is simpler and leaves less space for a possible mistake.

Interestingly, in our tests the running time of our solution in Picat 0.9 on small input is about 2.5 times larger than on the large input (table \ref{table:times}).
This is probably related to the implementation details and could change in the future versions.

\subsection*{Welcome to Code Jam\footnote{Problem link: \url{http://goo.gl/qeLls4}}}

``Welcome to Code Jam'' is a problem from the qualification round of GCJ 2009.
The task is to calculate the last 4 digits of the number of times the string ``welcome to code jam'' ($S$) appears as a subsequence of the given string ($T$).

This is a typical dynamic programming problem.
The problem state $dp[i][j]$ is the number of times the substring of $T$ of length $i$ contains the substring of S of length $j$ (modulo 10000).
The recurrence relation is: if $T[i] = S[j]$, $dp[i][j] = dp[i - 1][j - 1] + dp[i - 1][j]$, otherwise $dp[i][j] = dp[i - 1][j]$.

Our Picat program uses tabling \cite{warren1992memoing} (a kind of memoization) to implement the described dynamic programming solution in a top-down fashion.

\begin{lstlisting}[caption={Picat solution for the ``Welcome to Code Jam'' problem}]
s() = to_array("welcome to code jam").
table
ways(_, _, 0) = 1.
ways(_, 0, _) = 0.
ways(T, I, J) = W =>
    S = s(),
    if T[I] == S[J] then
        W = (ways(T, I - 1, J) + ways(T, I - 1, J - 1)) mod 10000
    else
        W = ways(T, I - 1, J)
    end.
do_case(Case_num, T) =>
    W = ways(T, length(T), length(s())),
    printf("Case #%w: %04d\n", Case_num, W).
\end{lstlisting}

Line 1 defines the string $S$ from the problem statement as a functional fact.
Lines 2--11 defines recursive \texttt{ways} function for dynamic programming. The calls to this function are automatically tabled (memoized) because of the \texttt{table} declaration.
Lines 3 and 4 describe base cases for the recursion. Lines 5--11 specify the recurrence relation.
The \texttt{do\_case} predicate in lines 12--14 calls the \texttt{ways} function and prints the results according to the problem specification.

An imperative solution can rely on the same recurrence relation, but might require more code to implement it either as a bottom-up dynamic programming or as a top-down recursion with memoization.

\subsection*{Bribe the Prisoners\footnote{Problem link: \url{http://goo.gl/pSbrTk}}}

Bribe the Prisoners was the hardest problem from the round 1C of GCJ 2009.
In it we have an array of $P$ prison cells, each cell is either empty or contains a prisoner. Every time a prisoner from one of the cells is released, all prisoners housed on either side of that cell until cell 1, cell P, or an empty cell get one coin each. Initially all cells contain prisoners.
Given a list of indices of prisoners to be released, find the minimum total number of coins that will be spent if the prisoners will be released in an optimal order.

This is an another dynamic programming problem. For each pair of cells $A \leq B$, $dp[A][B]$ is the best answer if prisoners occupy only cells from $A$ to $B$, inclusive.
If the first prisoner between $A$ and $B$ to be released is in cell $X$, $(B - A)$ coins are to be paid out immediately after his release, and then the smaller subproblems $dp[A][X-1]$ and $dp[X+1][B]$ have to be solved. The final answer $dp[1][P]$ corresponds to the initial state of all cells occupied.

Our Picat program uses mode-directed tabling \cite{zhou2010mode}.

\begin{lstlisting}[caption={Picat solution for the ``Bribe the Prisoners'' problem}]
table (+, +, +, min)
cost(A, B, FreeList, Cost) ?=>
    foreach(X in FreeList)
        (X < A ; X > B)
    end,
    Cost = 0.
cost(A, B, FreeList, Cost) ?=>
    member(X, FreeList),
    X >= A, X =< B,
    cost(A, X - 1, FreeList, CostLeft),
    cost(X + 1, B, FreeList, CostRight),
    Cost = B - A + CostLeft + CostRight.
do_case(Case_num, P, FreeList) =>
    cost(1, P, FreeList, Cost),
    printf("Case #%w: %w\n", Case_num, Cost).
\end{lstlisting}

The first line declares the tabling mode for the \texttt{cost} predicate: first 3 parameters are input parameters, and the last parameter is an output parameter
(the cost of releasing all prisoners in \texttt{FreeList} that occupy cells in the $[A; B]$ range) that must be minimized.
Lines 2--12 define two clauses of the \texttt{cost} predicate; both clauses are non-deterministic, and this is stated by using \texttt{?=>} syntax instead of \texttt{=>}.
The first clause states that if no prisoners in \texttt{FreeList} occupy cells between $A$ and $B$, the cost of their release will be 0.
The second clause calculates the release cost of prisoner X as described by the recurrence relation.
The \texttt{do\_case} predicate in lines 13--15 calls the \texttt{cost} function for the whole range of cells and prints the result according to the problem specification.

As with the previous problem, an imperative solution can use the same recurrence relation, but might require more code for a bottom-up or top-down approach implementation, including explicit comparison of release costs of different prisoners to find the minimum. Our Picat solution replaces most of the auxiliary code with a single \texttt{table} declaration.

\subsection*{Osmos\footnote{Problem link: \url{http://goo.gl/0N5zB8}}}

``Osmos'' is a problem from the round 1B of GCJ 2013.
The problem describes ``motes'' of different integer sizes. One mote (Armin) is controlled by a player, the rest are passive. If Armin is of size $X$, it can absorb any passive mote of size $Y < X$ and grow to size $X + Y$ as a result. You are given the initial size of Armin and the sizes of passive motes. You can add a passive mote of any positive size, or you can remove any existing passive mote. Minimize the number of addition and removal operations required for Armin to be able to absorb all passive motes.

Out Picat program uses the \texttt{planner} module \cite{zhou2014planning}.

To come up with an effective planning solution we need to notice that there always exists an optimal solution in which Armin absorbs passive motes in order from smallest to largest (if there is a pair of motes absorbed in different order, they can be swapped without increasing the number of operations needed).

\begin{lstlisting}[caption={Picat solution for the ``Osmos'' problem}]
import planner.
final([_, []]) => true.
action([Armin, Others], NewState, Action, Cost) ?=>
    Others = [Min | Rest],
    Armin > Min,
    NewArmin is Armin + Min,
    Action = absorb,
    Cost = 0,
    NewState = [NewArmin, Rest].
action([Armin, Others], NewState, Action, Cost) ?=>
    Others = [Min | _Rest],
    Armin =< Min,
    append(NewOthers, [_], Others),
    Action = remove,
    Cost = 1,
    NewState = [Armin, NewOthers].
action([Armin, Others], NewState, Action, Cost) ?=>
    Others = [Min | _Rest],
    Armin =< Min,
    NewItem is Armin - 1,
    NewOthers = [NewItem | Others],
    Action = add,
    Cost = 1,
    NewState = [Armin, NewOthers].
do_case(Case_num, Armin, Others) =>
    Limit = length(Others),
    best_plan([Armin, sort(Others)], Limit, _Plan, Cost),
    printf("Case #%w: %w\n", Case_num, Cost).
\end{lstlisting}

Solving a planning problem in Picat requires a \texttt{final} predicate and an \texttt{action} predicate.
Line 2 defines the \texttt{final} predicate which has one parameter -- the current state -- and succeeds if the state is final.
In our program a state is represented as a 2-element list: the first item is the Armin size, and the second item is a sorted list of the sizes of passive motes (\texttt{Others}). A state is final if the \texttt{Others} list is empty.

Lines 3--24 define the \texttt{action} predicate which has three clauses -- one for \texttt{absorb}, \texttt{remove} and \texttt{add} actions -- and has four parameters: current state, new state, action name, and action cost.
Lines 3--9 define the \texttt{absorb} action which can be used if Armin is bigger than the first of the other motes at the cost of 0.
Lines 10--16 define the \texttt{remove} action which removes the last (the largest) mote from \texttt{Others} at the cost of 1.
The \texttt{append} predicate and the \texttt{[ | ]} syntax for getting the head and the tail of a list work exactly the same way as in Prolog.
Lines 17--24 define the \texttt{add} action which adds a mote of size $Armin - 1$ to the beginning of the \texttt{Others}
(so it can be absorbed by the next absorb action) at the cost of 1.

Picat's predicate for finding an optimal plan -- \texttt{best\_plan} -- has two input parameters: the initial state and the resource limit, and two output parameters: the best plan and its cost. To find an optimal plan the system uses tabling and iterative deepening depth-first search-like algorithm. If no plan was found and the maximum resource limit was reached, the predicate fails.
In this problem the resource limit for \texttt{best\_plan} is the initial number of passive motes, because there is an obvious plan of this cost to remove all the motes.

This solution is a declarative specification of the problem statement which relies on just a few observations about the problem.
An imperative solution would require much more insight into the problem. One could notice that in an optimal solution if a mote is removed, all motes of equal or greater sizes are also removed (if one of larger motes is absorbed, so can be the mote in question). Thus, a greedy solution is: keep absorbing passive motes from smallest to largest while absorbing the next one is possible. After this, either remove all passive motes left or keep adding motes of size one less than Armin's current size and immediately absorbing them until Armin can absorb the smallest passive mote left. Repeat until Armin absorbs the last of the given passive motes.

\begin{table}
\centering
\caption{Running times for small (4 minutes time limit) and large (8 minutes) inputs\protect\footnotemark}
\label{table:times}
\begin{tabular}{llrr}
\hline
Problem               & Technique &  Small   & Large \\
\hline
Triangle Areas        & constraint programming      & 2.4s & 0.9s \\
Welcome to Code Jam   & dynamic programming         & 0.0s & 0.3s \\
Bribe the Prisoners   & dynamic programming         & 0.0s & 4.7s \\
Osmos                 & planning                    & 0.0s & 0.1s \\
\hline
\end{tabular}
\end{table}
\section{Conclusions}

We have given several examples of declarative solutions for GCJ problems with Picat using constraint logic programming and tabled logic programming.

We considered using Picat's mixed integer programming module which might be useful for solving many GCJ problems \cite{gcj-eclipse-arxiv},
but currently there is no easy way to suppress log messages written to standard output by the underlying solver.

Compared to Prolog, Picat code can be more compact because of functions (function calls can be nested,
so there is no need for intermediate variables), list comprehensions, and more convenient console input/output.
Also, while many modern Prolog systems have loop constructs,
Picat loop syntax looks much cleaner because neither global nor local variables need to be explicitly declared.

Running times of our Picat programs are several orders of magnitude smaller than the time limit imposed by GCJ rules (table \ref{table:times}).

We also have found that GCJ problems can be complex and large enough to exercise many different aspects of a programming language implementation:
we discovered and reported two serious bugs in the version 0.8 of the Picat system while working on this paper (the bugs were promptly fixed for Picat 0.9).

\footnotetext{Results were obtained on a 64-bit Linux machine with Intel Core i7-4900MQ CPU @ 2.80GHz and 16GB RAM using Picat 0.9. }

\bibliographystyle{abbrv}
\bibliography{gcj-picat}

\end{document}